# Quantum Image Processing - Challenges and Future Research Issues

[1]Sanjay Chakraborty, [2]Sudhindu Bikash Mandal and [3]Soharab Hossain Shaikh

[1,2]*A.K.Choudhury School of Information Technology, University of Calcutta, Kolkata, India*
schakraborty770@gmail.com
sudhindu.mandal@gmail.com

[3]*Computer Science and Engineering, BML Munjal University, India*
soharab.hossain@gmail.com

**Abstract**
*Image processing on quantum platform is a hot topic for researchers now a day. Inspired from the idea of quantum physics, researchers are trying to shift their focus from classical image processing towards quantum image processing.This paper starts with a brief review of the principles which underlie quantum computing, and also deals with some of the basics of qubits and quantum computing.Then this paper starts to deal with some different methods of image storage, representation and retrieval in a quantum system.This paper also describes the advantages of using those methods in quantum systems compare to classical systems. In the next section, a short discussion on some of the important aspects, comparison among them and applications of quantum image processing is presented. A few other hot topics and open problems in quantum image processing are mentioned in this paper. This review article will provide the readership an overview of progress witnessed in the area of Quantum Image processing while also simulating further interest to pursue more advanced research in it.*

## 1. Introduction

Where the concept of 'Classical computation' ends, there the concept of 'Quantum computation' is started. Quantum computing is an extended part of quantum physics where the main skeleton is made of by quantum mechanics. Quantum physics and quantum mechanics together reveal several facts of our universe regarding elementary atomic structure. Combining these very same concepts with computing, quantum computing tries to improve the overall performance of computation. By exploiting the laws of quantum physics,researchers are getting great performance improvements in different areas like,information encoding, processing and communication [3][20].

### 1.1 Why Quantum Computing?

Quantum computing is an emerging technology. Moore's law states that computer power will double for constant cost roughly once every two years [2]. To find a possible solution of this problem quantum computation concept has been introduced. This concept will help us to build much smaller embedded electronic devices to perform computation more quickly, effectively and efficiently [6]. According to the principles of quantum physics and quantum mechanics, the computing power of a quantum machine is enormous compared to that of a classical one[13]. Quantum computing is a part of futuristic computing but experiments have been carried out in which quantum computational operations were executed on a very small number of quantum bits. And also several government organizations and agencies are funding quantum computing research in order to develop quantum computers for medical imaging, civilian, business, trade, education & research and some security purposes. The various applications of quantum computing is described in the Figure 1.

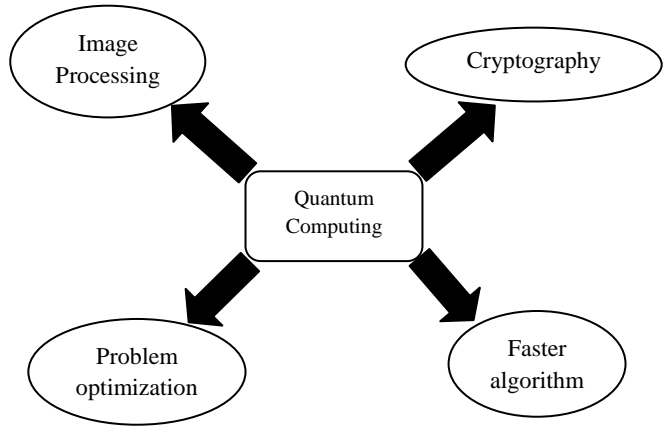

Fig.1 Applications of Quantum Computing

When these futuristic machines will be built in a commercial way, then it will definitely reduce the effort, time, size and space problems from various trends. Encouraged by this idea, the applications of quantum computation are spread over several areas which is represented by figure1. Among all these fields, quantum computing has an important impact on image processing field. Now a day, many researchers are working on quantum image processing field in order to optimize the various image processing algorithms in terms of their quickness, efficiency and accuracy. This article mainly deals with the implication of quantum computing in the field of image processing. This article also deals with the retrieval, processing, storing, enhancement, compression and restoration of the visual information (2D image data). Major questions that we are mainly focus to describe throughout this paper are as follows,

- What is the advantage of using quantum computing approach in image processing?
- What is the difference between digital and quantum images?
- How representation, storing and retrieval of 2D images manage into quantum environment?
- How we can store images through quantum entanglement?
- How quantum algorithms make an effect to construct Image Histogram?
- How we can apply noise filteration on an image in a quantum computer?
- How image compression is achieved into a quantum environment?

## 1.2 What is Qubit in Quantum Computing?

The basis of classical computing is bits whereas the basis of quantum computing is qubits. Quantum computation and quantum information are built upon an analogous concept of qubits. Qubit represents some sort of atomic structure like, hydrogen atom. It is confined to be $|0\rangle$ and $|1\rangle$ states which represents ground and excited states respectively [7]. The pictorial representation is enclosed in Figure 2. The general state of an electron can be represented by the superposition ($|S\rangle$) of the above two qubits states,

$$|S\rangle = a|0\rangle + b|1\rangle \qquad (1)$$

Where a and b belongs to a complex vector space and $|a|^2 + |b|^2 = 1$, which represents square of the length of the vectors, sometime called "Unit Vector". They are also called probability amplitudes. So now we can define that qubit is unit vector in a 2D complex vector space ($|\rangle$ -Dirac's Ket notation)[11] [32].

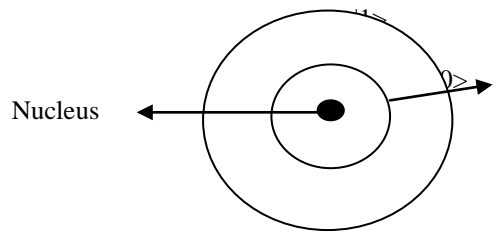

Fig.2 Orbital representation of an atom through qubits

If it maps on 2D complex vector space then it can be represented in column vector form like, $\begin{pmatrix}a\\b\end{pmatrix}$. So for ground state $|0\rangle$, it is $\begin{pmatrix}1\\0\end{pmatrix}$ where, a=1 & b=0 and for excited state $|1\rangle$, it is $\begin{pmatrix}0\\1\end{pmatrix}$ where, a=0 & b=1.

### 1.2.1 Qubit measurements and Unit circle theory

Let just draw two states $|0\rangle$ and $|1\rangle$ respectively with their basis vectors $[1\ 0]^T$ and $[0\ 1]^T$. In figure3, $|0\rangle$ is represented according to X-axis and $|1\rangle$ is represented according to Y-axis. Let draw some other vectors with two different angles. The first vector whose angle is 45 degree with X-axis represent the basis vector $[1/\sqrt{2}\ \ 1/\sqrt{2}]^T$ and an another vector whose angle is 60 degree with X-axis represent the basis vector $[1/2\ \ \sqrt{3}/2]^T$.

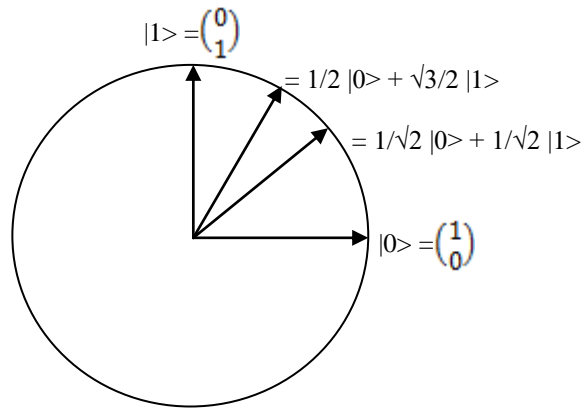

Fig.3 Unit Circle representation on various angels

So from here we learn that a qubit is a unit vector in 2D complex vector space. So if it represents in the form of "Bloch sphere", then |0> represents X-axis and |1> represents Y-axis of the sphere respectively whereas any point over the surface of the sphere represents a superposition bit of |0> and |1> or a unit basis vector[7]. There are two types of measurements, one measurement based on the standard basis states and second, measurement based on the arbitrary basis states.

**1.2.1.1 Measurement on standard basis states:**

Suppose we have a state |S> which makes an angle θ with |0> state (X-axis). Obviously, all amplitudes are real as it has been drawn in a 2D real space (figure 4)[7].

$$|S> = \cos\theta\ |0> + \sin\theta\ |1> = \begin{pmatrix}\cos\theta\\ \sin\theta\end{pmatrix} \quad (2)$$

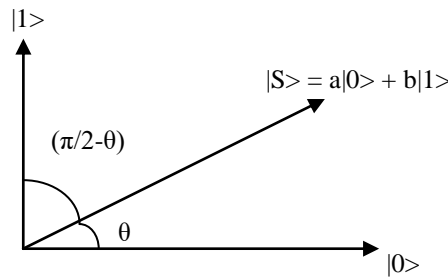

Fig.4 Projection on to standard basis states

From the figure 4, |0> with probability $\cos^2\theta$ and |1> with probability $\sin^2\theta = \cos^2(\pi/2-\theta)$ So the state |S> is projected either onto ground state or excited state with the above probabilities.

**1.2.1.2 Measurement based on the arbitrary basis states:**

Here instead of measuring on to |0> and |1> basis, we can measure it in any orthogonal basis of our choice[7]. Let consider the below figure,
In Figure5, state |S> is measured with respect to |u> and |u'> basis, where |u> is measured with probability $\cos^2\theta$ and |u'> is measured with probability $\sin^2\theta$. So the amplitude value will be,

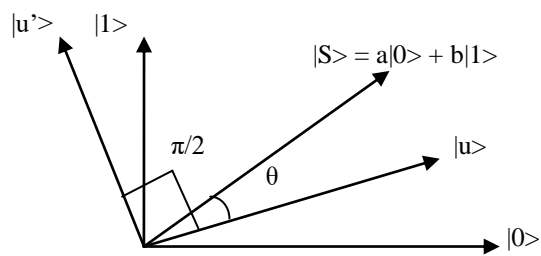

Fig.5 Projection on to arbitrary orthogonal basis states

$|u\rangle = 1/\sqrt{2}\ |0\rangle + 1/\sqrt{2}\ |1\rangle$

$|u'\rangle = -1/\sqrt{2}\ |0\rangle + 1/\sqrt{2}\ |1\rangle$

## 1.3 What is Quantum Image Processing (QIMP)?

Images are one of the major sources of information transmitted and stored in the digital media. They are widely used in satellite communication, social media, medical diagnosis and other applications in our daily life where visual content is required. Digital image processing refers to processing digital images by means of a digital computer whereas quantum image processing refers to processing images by means of a quantum computer[8][20]. Quantum Image Processing is a subset of Quantum Information Processing. Due to the huge development of quantum computation and quantum information, it may be good ways of understanding the behaviour of visual information. Quantum Image Processing shows how to store and represent an image in a quantum system using the unique properties of quantum mechanics such as, superposition, parallelism, cohesion and entanglement[10][20]. So in one sentence we can say that it is a collection of quantum based methods to speed up the processing time with efficiency of various image processing algorithms. Impact of quantum methods over image processing is shown in Figure 6.

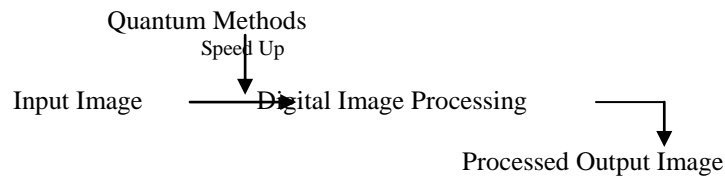

Fig.6 Flowchart of Quantum Image processing (QIMP)

## 2. Quantum Image Processing

Image processing such as image storage, retrieval, representation, processing, noise filteration, compression etc. are given better results and approach in quantum computing environment rather than any classical computing environment. This paper mainly deals with many different techniques developed till now to handle the processing of images efficiently in quantum computing environment. Those various approaches are described briefly in this section.

### 2.1 Quantum image processing (QIMP) vs. Classical image processing
Some crucial points to highlight the differences between Quantum image processing and Classical image processing are discussed below,

- In quantum computing, the continuous nature of the qubit helps us to store image information (color or position values) without having to pre-process it. Here the actual color value of a pixel is analyzed by reading its corresponding frequency[1].
- Some problems like white noise and the difficulty of reaching exquisite levels of control are always present in classical system.
- As all of we know that there is a tight dependency from one part of an image to other parts. Maintaining this correlations between different points of an image are very important in order to properly understand and describe the image. Storing of an image in classical computers may lose such relevant information, because it needs information about the relationship between stored color values (or gray-scale) and corresponding pixels locations. The advantage of the quantum computers lie here[1]. Quantum computer can store color values (or gray-scale) and corresponding pixels locations of an image using the concept of quantum registers.
- The complexity of QIMP algorithms is usually computed in terms of quantum gates, while classical complexity is usually computed in terms of algorithm running time. Quantum hardware designers must clearly identify the amount of classical and quantum resources needed to implement a given QIMP protocol.

### 2.2 Storing, Retrieval and Representation of an image in quantum computing environment

**Quantum Normalized Amplitude Based Image Processing (QNAIMP):**
As above said, Image can be represented on quantum computers based on the location of pixel values and the normalized amplitude of the signal values of a gray scale or a color image. The proposed representation focuses on two features of images, the amplitude of the image signal and its corresponding pixel location on the image space. In this approach an image is represented by two-dimensional matrix of row and column vectors. Each pixel location in a classical image is specified by a row and a column number (like, f(x,y), where f represents the amplitude value of the pixel and (x,y) represents the pixel's positions). Similarly, any pixel can also be defined by row and column vectors on a quantum computer. According to one famous approach based on normalized amplitude, suppose we have M-length row location vector with m qubits and N-length column location vector with n qubits [4][5]. The recursive tensor products of the various permutations of these constituent vectors form M-length row-location vector using m-qubits. Mathematically, it is defined as,

$$|I\rangle_p = |i\rangle^{\otimes m}$$

Where, $|I\rangle_p$ is the row-location vector or the state of m-qubit, $i \in \{0, 1\}$, $m = \log_2 M$ and p is the row number of pixel. Likewise, orthonormal constituent vectors for N-length column-location vector can be written as,

$$\langle 0| = (1\ 0)$$
$$\langle 1| = (0\ 1)$$

This '$\langle\ |$' is called bra notation. The recursive tensor products of the constituent vectors using bra notation is,

$$\langle J|_q = \langle j|^{\otimes n} \tag{3}$$

Where, $\langle J|_q$ is the column-location vector or the state of n-qubit, $j \in \{0, 1\}$, $n = \log_2 N$ and q is the row number of pixel. To represent a pixel location in its complete two-dimensional format, the outer product of the row-location vector and the column-location vector is taken in Table 1. Hence, the two-dimensional location of a pixel can be identified by the following,

$$L_{pq} = |I\rangle_p \otimes \langle J|_q \tag{4}$$

Where, $L_{p,q}$ is the pixel at pth row and qth column.

Table1. Two dimensional Quantum state of each pixel location of a 4 × 4 size image matrix

|   | 1 | 2 | 3 | 4 |
|---|---|---|---|---|
| 1 | $\|00\rangle\otimes\langle00\|$ | $\|00\rangle\otimes\langle01\|$ | $\|00\rangle\otimes\langle10\|$ | $\|00\rangle\otimes\langle11\|$ |
| 2 | $\|01\rangle\otimes\langle00\|$ | $\|01\rangle\otimes\langle01\|$ | $\|01\rangle\otimes\langle10\|$ | $\|01\rangle\otimes\langle11\|$ |
| 3 | $\|10\rangle\otimes\langle00\|$ | $\|10\rangle\otimes\langle01\|$ | $\|10\rangle\otimes\langle10\|$ | $\|10\rangle\otimes\langle11\|$ |
| 4 | $\|11\rangle\otimes\langle00\|$ | $\|11\rangle\otimes\langle01\|$ | $\|11\rangle\otimes\langle10\|$ | $\|11\rangle\otimes\langle11\|$ |

Now, the normalized amplitude image matrix (α) can be defined as,

$$\alpha_{p,q} = \sqrt{AMP_{p,q}} / \sqrt{\sum_{p=1}^{M}\sum_{q=1}^{N} AMP_{p,q}} \tag{5}$$

Where, $AMP_{p,q}$ be the amplitude of the digital image signal at pth row and qth column, where $p \in [1,M]$ and $q \in [1,N]$. After achieving the normalized amplitudes of the image signal in the above equation, the next task is to represent these amplitudes also by multiple qubits quantum states. The reason for this approach is to reduce the normalized amplitude values to quantum states so that the amplitudes can easily be stored and transmitted in a quantum environment. Now we can replace $\alpha_{p,q}$ by the following,

$$\alpha_{p,q} = \alpha^{\eta}_{p,q} / \alpha_D \tag{6}$$

Where, $\alpha^{\eta}_{p,q}$ and $\alpha_D$ are real and are represented by qubits As follows,

$$|\alpha^{\eta}_{p,q}\rangle = |i\rangle^{\otimes \log_2 \alpha^{\eta}_{p,q}}$$

$$|\alpha_D\rangle = |i\rangle^{\otimes \log_2 \alpha_D}$$

The following equation defines the proposed quantum image representation mathematically,

$$|Z\rangle = \sum_{P=1}^{M}\sum_{q=1}^{N} \alpha^{\eta}_{p,q} \times L_{pq} \tag{7}$$

Where $|Z\rangle$ is a two-dimensional gray scale quantum image or one of the color channels of the quantum color image. In this approach, instead of storing and transmitting amplitudes of the image signal, what actually stored and transmitted are all the pixel locations having the same amplitude values, for all the amplitudes present in the image [4].Paper [1] [12] and [13] follows the same approach on how a color can be stored in a qubit of a quantum computer system. A digital image comprises of a set of pixels containing the intensity values. The intensity value of a pixel determines its colour. The Colour can be represented and analysed by its frequency. From Figure 7, it can be shown that a quantum machine can be used to detect electromagnetic waves of different frequencies and maps them into corresponding qubits. We can measure the superposition of the quantum state as,

$$|S\rangle = \cos\theta/2 |0\rangle + \sin\theta/2 |1\rangle \qquad (8)$$

Where,
$$\cos^2\theta/2 + \sin^2\theta/2 = 1$$

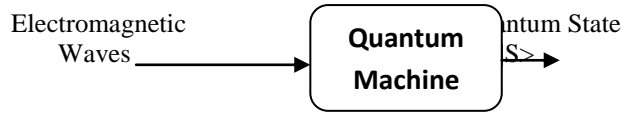

Fig.7 Electromagnetic Waves to Quantum state

Here the parameter θ varies with the frequency of the colour. Each value of a θ determines the state of a qubit. These qubits can be stored in a qubit lattice Q in a quantum memory. The pictorial representation is shown in Figure 8. This Q is represented in 2-dimensional qubit array as a form,

$$Q = \{|q\rangle i,j\}, i\epsilon\{1, 2, \ldots, n_1\}, j\epsilon\{1, 2, \ldots, n_2\}$$

The goal is to store visual information in Z where Z is a 3-dimensional set of qubit lattices and each layer $Q_k \epsilon Z$ will be used to store a copy of the image.

| $|S\rangle_{11}$ | $|S\rangle_{12}$ | ................. | $|S\rangle_{1m}$ |
|---|---|---|---|
| $|S\rangle_{21}$ | $|S\rangle_{22}$ | | $|S\rangle_{2m}$ |
| ...... | ..... | ................. | ..... |
| $|S\rangle_{n1}$ | $|S\rangle_{n2}$ | | $|S\rangle_{nm}$ |

Fig.8 Graphical representation of a set of qubits in $Q_1$ lattice

Image retrieval is a process of retrieving stored images from the database. For an efficient image retrieval process a quantum mechanical approach is followed by paper [12]. According to this approach suppose a 2X2 color image is stored in a qubit lattice Q. Suppose $|S_{ij}\rangle$ are the quantum states of colors $C_1, C_2, C_3, C_4$ stored in the qubit lattice. The image contains four different colors out of which colors $C_1$ and $C_3$ have states $|0\rangle$ and $|1\rangle$ respectively. The color probability can be measured by preparing two qubit maximally entangled state, one state measurement will be done for the opposite value of spin and other state measurement will be done for the same value of spin. It is presented by Figure 9.

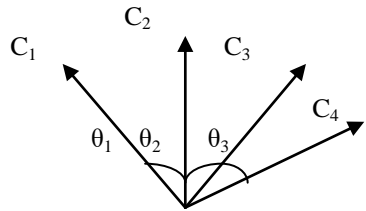

Fig.9 Color quantum states measurement using angular values

With this strategy the expression for their probabilistic measurements is given by

$$1/\sqrt{4}[\cos^2(\theta_1/2) + \cos^2((\theta_1+\theta_2)/2) + \sin^2(\theta_3/2) + \cos^2((\theta_3+\theta_2)/2)] \qquad (9)$$

Maximizing the above expression with respect to $\theta_1, \theta_2$ and $\theta_3$, we get $\text{Prob}(C_2) = \cos^2(\pi/8)$ and $\text{Prob}(C_4) = \cos^2((\pi/4 - \pi/2)/2)$ [12].

**Quantum Entangled Image Processing (QEIP):**
According to paper[5][22][39], we have a new method to store an image in a quantum system by using the concept of quantum entanglement. Quantum entanglement is a kind of correlation which is used to store binary images in this approach. Here first consider an array of n qubits where each qubit associated with two parameters x and y, which together represent grid points of some simple 2D image. This array has been stored in a memory register.

$$|S_{initial}\rangle = \bigotimes_{i=1}^{n} |0\rangle_{i(x,y)} \qquad (10)$$

A white point is associated on the grid with qubit state |0>, whilst black corresponds to state |1>. However certain classical approaches are extended due to fully exploit the unique properties of Entanglement. Here a object shape, as for example shown in Figure 10, a square shape is needed to be stored in our n qubit array by choosing each vertex of the square and its corresponding qubit to |1>.

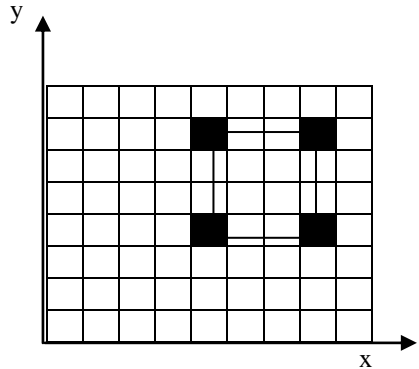

Fig.10 Simple storage procedure for a single square in a qubit array

However, the use of Entanglement between vertex locations (straight lines) provides a more fruitful approach. The appropriate vertex positions may then be retrieved by applying Grover's quantum search algorithm to the array. For four vertices stored in the array, application of Grover's search algorithm will require approximately √n/4 steps to recuperate the information specifying the locations of the vertices of the square. The image of the above square is then very simply reconstructed from this information [5].

Now we are moving towards some different approach of representing some basic image processing operations by implementing through quantum circuit [13]. Here image negative can be achieved by quantum NOT gate. It can be achieved with the following transform:

$$s = T(r) = L - 1 - r;$$

Where, r and s are the pixel values before and after processing, respectively. The quantum analogous of the classical NOT (CNOT) gate is labelled X and can be defined such that X|0> = |1> and X|1> = |0>. The quantum NOT gate acts similarly with its classical counterpart, although, unlike in the classical case, its action is linear: state a|0>+b|1> is transformed in a corresponding state b|0>+a|1>. A convenient way of representing the action of the quantum NOT gate is shown in Figure 11 and its corresponding matrix form is,

$$X = \begin{pmatrix} 0 & 1 \\ 1 & 0 \end{pmatrix}$$

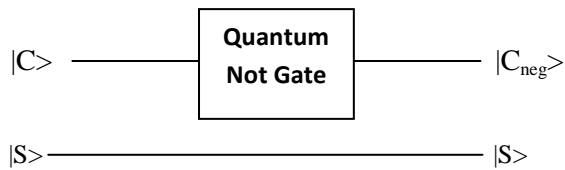

Fig. 11 Quantum circuit implementing the image negative

Where, |S> is a superposition state. Quantum image binarization and quantum image histogram analysis are also presented through quantum circuit representation in this paper[13]. They mainly used the concept of if-then-else or quantum comparator to represent the threshold based image segmentation(also called image binarization).The latter operations are built using a special quantum bit comparator, CNOT gates and multiple control Toffoli gates [13]. A very new threshold based quantum color image segmentation technique has been introduced in the paper [56]. This article has been showed the multilevel color image segmentation circuit design based on some predefined threshold using the concepts of quantum parallelism, superposition and a basic quantum oracle [56].

In order to perform image retrieval, content based retrieval is one of the famous techniques in image processing. Content Based Image Retrieval (CBIR) is a technique in which images are indexed based on their visual contents and retrieving is only based upon these indexed images contents. Visual content means the shape, color, texture etc. of an image object. This approach introduces a new easy shape representation for content based image retrieval technique by deriving the concept of quantum superposition with the basis of distance histogram[19].Here the concept of feature vector is introduced. It represents image content as numeric values.Then these numeric values are used to derive for the whole shape boundary of the image. This technique is called as contour based shape representation method using shape signature.In this approach,two shapes are matched by finding the distance between their corresponding feature vectors which is usually led by Euclidian distance or city block distance[19].Most shape signatures are quantized into signature histogram which is rotationally invariant. Centroidal distances histogram is an actual contour shape representation method which is invariable to translation and rotation. Normalization makes distances histogram inflexible to scaling.In feature extraction phase,the centroidal distances are computed between points on the shape boundary and the centroid of shape, these distances contain information about the shape of target which used distance histogram (DH) to describe the shape of target by sampling the centroidal distances into buckets resulting in a histogram denoted as:

$$D = (d_0, d_1, \ldots, d_{N-1})$$

Where N is the number of buckets in the histogram and $d_i$ is the total number of centroidal distances which were discretized into bucket $i$.

The ($m,n$) pixel qubit of an normalized image f(m,n)can be defined as:

$$|f(m,n)\rangle = \sqrt{1 - f(m,n)} |0\rangle + \sqrt{f(m,n)} |1\rangle \qquad (11)$$

Such that $\sqrt{1 - f(m,n)}$ and $\sqrt{f(m,n)}$ are the appearance probabilities for pixel grey level being at state 0 and 1 simultaneously. The proposed method combines distance histogram (DH) and quantum superposition in this approach. This combined approach is called Quantum distance histogram(QDH). After transforming the centroidal distances quantum space, then a measurement is performed per each point to find its state. The transformation of point distance d(x,y) into quantum bit d(x,y) is the superposition of two quantum states |0> and |1> which is expressed by the following equation:

$$|d(x,y)\rangle = \sqrt{1 - d(x,y)} |0\rangle + \sqrt{d(x,y)} |1\rangle \qquad (12)$$

For certain point centroidal distance $d_i(x,y)$, its correlative characteristic with neighboring pre-post points centroidal distances along shape boundary is made up by superpositioning their state in bi-qubits system defined as:

$$|d_{i-1}(x,y), d_{i+1}(x,y)\rangle = \sqrt{1 - d_{i-1}(x,y)} \sqrt{1 - d_{i+1}(x,y)} |00\rangle + \sqrt{1 - d_{i-1}(x,y)} \sqrt{d_{i+1}(x,y)} |01\rangle + \sqrt{d_{i-1}(x,y)} \sqrt{1 - d_{i+1}(x,y)} |10\rangle + \sqrt{d_{i-1}(x,y)} \sqrt{d_{i+1}(x,y)} |11\rangle \qquad (13)$$

Thus, the state of point location oscillates among |00>, |01>, |10> and |11> states until it breakdown into definite state by measurement, thereby, existing this action unveiled from the quantum level [19].

A Flexible Representation of Quantum Images (FRQI) is proposed to provide a representation for images on quantum computers in the form of a normalized state which captures information about colors and their corresponding positions in the images [24]. It introduces an algorithm for quantum image compression (QIC), and processing operations for quantum images are combined to build the whole process for quantum image processing on FRQI. The whole procedure is divided into three steps,

**The preparation for FRQI:** The preparation process for FRQI is indicated by using Hadamard and controlled rotation operations. A Polynomial Preparation theorem is used here to represent that the total number of simple operations used in the process is polynomial for the number of qubits which are used to encode all positions in an image. Here a representation for images on quantum computers can be done by capturing information about colors and their corresponding positions. It can be represented as [37],

$$|I(\theta)\rangle = 1/2^n \sum_{i=0}^{2^{2n}-1} (\cos\theta_i |0\rangle + \sin\theta_i |1\rangle) \otimes |i\rangle \qquad (14)$$

Capturing information about colors and the corresponding positions of those colors, where $\otimes$ is the tensor product notation, $|0\rangle$, $|1\rangle$ are 2-D computational basis quantum states $|i\rangle$, i=0, 1, 2…., $2^{2n}-1$ are $2^{2n}$ -D computational basis quantum states and $\theta$ is the vector of angles encoding colors. $\cos\theta_i|0\rangle + \sin\theta_i|1\rangle$, this encode the information about colors and $|i\rangle$ is the corresponding position in the image [18][33]. The FRQI state is normalized as, $\| |I(\theta)\rangle \| = 1$. For 2×2 image, the FRQI state is,

$$|I(\theta)\rangle = \frac{1}{2} [(\cos\theta_0|0\rangle + \sin\theta_0|1\rangle)\otimes|00\rangle + (\cos\theta_1|0\rangle + \sin\theta_1|1\rangle)\otimes|01\rangle + (\cos\theta_2|0\rangle + \sin\theta_2|1\rangle)\otimes|10\rangle + (\cos\theta_3|0\rangle + \sin\theta_3|1\rangle)\otimes|11\rangle] \qquad (15)$$

Where $\theta_0, \theta_1, \theta_2$ and $\theta_3$ are four different vectors of angles encoding colors[21][24][27].

**Invertible image processing operators on FRQI:** With the FRQI proposal, images are expressed in their FRQI states and quantum image processing operations are performed using unitary transforms on those states. This transform is divided into three types of operators. Operators in the first group use only information about the color, such as color shifting and the second group contains those based on colors at some position in the images, for instance the changes in color at specific positions. The last group targets information about both color and position as in Fourier transform. Each category has its own type of unitary transform [24][40].

Upto this whatever image representation techniques discussed are not suitable for constructing image histogram for different gray scales images. According to paper [14], it proposes a novel image representation model using a quantum statistical algorithm for constructing image histogram. It is designed in such a way that it can achieve an approximately quadratic speedup over the classical constructing method. This approach is proposed to remove the deficiency of FRQI approach of paper[24]. Image histogram acts as a graphical representation of the grayscale distribution in a digital image. According to this paper[14], two entangled quantum registers are utilized to store the grayscale qubit sequence and the position qubit sequence, respectively. Via quantum counting algorithm, the new proposed quantum histogram construction can achieve an approximate quadratic speedup over the classical counterpart [14].

There are several drawbacks associated with FRQI models,
- Due to the probability amplitudes of a quantum state cannot be accurately determined using a finite number of measurements; original classical image cannot be retrieved.
- In FRQI model, only a single qubit is used to represent the color information for each pixel, that's why it is difficult to design complex quantum circuits for images where separate pixels with different colors needs to be applied (like, Image Histogram).

So according to the analysis, the model is suitable for grayscale images but it may be unfeasible to represent true color images ($2^{24}$ bits). But one modified approach is introduced by Caraiman and Manta[31].

**Caraiman's Quantum Image Representation Model (CQIR):**
In this model, the color and position information of a pixel can be retrieved in a finite deterministic way rather than probabilistic measurements (FRQI). This approach contains at most $2^m$ possible colors for an image and the color of a pixel with position i is expressed by means of a superposition of all $2^m$ possible colors [31]. So, the color $C_i$ of pixel i can be expressed as,

$$|C_i\rangle = \sum_{j=0}^{2^m-1} \alpha_{ij} |j\rangle \quad (16)$$

Where $C_i$ denotes the color of a pixel i, the image contains at most $L=2^m$ colors, and the coefficients $\alpha_{ij}$ are used to express the color of a pixel with position *i* by means of a superposition of all *L* possible colors [35]. Then the register $|P\rangle$ encodes pixel positions using 2*n* qubits and the quantum image is represented using an (m+2n)-qubit register that encodes both the color and position of the pixels,

$$|Q\rangle = |C\rangle_m \otimes |P\rangle_{2n} \quad (17)$$

As for example, a 2X2 quantum image with four possible colors can be represented as,

$$|Q\rangle = 1/\sqrt{2^2}(|01\rangle|00\rangle+|00\rangle|01\rangle+|11\rangle|10\rangle+|10\rangle|11\rangle)$$

Where, $|01\rangle, |00\rangle, |11\rangle$ and $|10\rangle$ are the color qubits of first, second, third and fourth co-ordinate of the 2X2 image respectively. In this approach, both the color and position of a pixel can be retrieved deterministically through a finite number of projective measurements. A larger class of more complex image processing operations can be applied using this model. Compared to the FRQI color representation mode, this approach requires *m* qubits to represent $L=2^m$ color values instead of one color qubit [31][36].

Similar to FRQI model, there is another model which encodes the color information in the probability amplitudes of the one qubit state. This probability amplitude can be measured by the bijective relationship between the frequency of the monochromatic electromagnetic wave (describes the color) and the angle parameter of a qubit. The same model is applied to infrared images where the injective function is used to measure probability amplitude of the corresponding one qubit state [38]. In these above models, there are some drawbacks which lead to a set of problems to represent original color image accurately. The above approaches are bounded by the practical limitations of acquiring probability amplitudes for qubit state after measuring the energy dissipation of the electromagnetic wave (describes the color) and the angle parameter. If the $\alpha_{ij}$ in the equation (16) becomes real number, then a statistical procedure is required to retrieve the pixel color. It is a time consuming and expensive procedure. The above model which requires one qubit to represent color information of a pixel is superior than the model of the paper [31], requires m qubits to represent $2^m$ color values. But both models are not using the basic simple quantum gates and multi-valued quantum computing logic which leads to the simple representation of specially RGB color images. Such a multilevel based approach would allow for a significant decrease in the state space needed to represent quantum images while providing the means for faster and more efficient operations[54].

**Multi-Channel Quantum Image (MCQI) Representation:**
MCQI technique is represented with the modified approach of FRQI technique. MCQI is actually a color representation of FRQI approach [49][50]. As the name suggest, MCQI uses R, G and B channels to represents various color information of an image and holds the normalized state. In this approach, three qubits are used to encode color information of an image. This can be represented as,

$$|I\rangle = 1/2^{n+1} \sum_{i=0}^{2^{2n}-1} |C^i_{RGB\alpha}\rangle \otimes |i\rangle \qquad (18)$$

Where the color information $|C^i_{RGB\alpha}\rangle$ encoding the RGB channels information is defined as:

$$|C^i_{RGB\alpha}\rangle = \cos\theta^i_R |000\rangle + \cos\theta^i_G |001\rangle + \cos\theta^i_B |010\rangle + \cos\theta_\alpha |011\rangle + \cos\theta^i_R |100\rangle + \cos\theta^i_G |101\rangle + \cos\theta^i_B |110\rangle + \sin\theta_\alpha |111\rangle, \qquad (19)$$

Where, $\{\theta^i_R, \theta^i_G, \theta^i_B\} \in [0,\pi/2]$ are three angles encoding the colors of the R, G, and B channels of the $i^{th}$ pixel, respectively and $\theta_\alpha$ is set as 0 to make the two coefficients constant ($\cos\theta_\alpha = 1$ and $\sin\theta_\alpha = 0$) to carry no information. MCQI technique encode R, G and B channel information using fewer bits due to the utilization of the property of quantum parallelism to encode color information and their corresponding positions. In terms of fewer bits requirement, MCQI plays better role that qubit lattice based approach [51].

**Novel enhanced quantum representation (NEQR) [42]:**

Unlike FRQI model, a novel enhanced quantum representation (NEQR) technique stores the grayscale values of every pixel of an image using the basis states of a qubit sequence. It is not encoded the color values of pixels through the angle parameters. To store grayscale values and positional information of all the pixels of an image, NEQR uses two entangled qubit sequences. In this approach, the gray range of an image is represented with a binary sequence which encode the grayscale value f(x,y) as,

$$f(x,y) = \bigotimes_{i=1}^{q} |C^i_{xy}\rangle \qquad (20)$$

Where, the gray range of an image is $2^q$. The representative expression of a $2^n \times 2^n$ version of such a quantum image can be written as:

$$|I\rangle = 1/2^n \sum_{x=0}^{2^n-1} \sum_{y=0}^{2^n-1} |f(x,y)\rangle |xy\rangle$$

$$= 1/2^n \sum_{x=0}^{2^n-1} \sum_{y=0}^{2^n-1} \bigotimes_{i=1}^{q} |C^i_{xy}\rangle \qquad (21)$$

In NEQR model, it utilizes the basis state of qubit sequence to represent the grayscale of pixels instead of probability amplitude of a single qubit used in the FRQI model. Therefore, the time complexity of preparing the NEQR quantum image exhibits an approximately quadratic decrease, i.e. $O(qn.2^{2n})$, compared to FRQI approach. Besides that, the image compression (using minimization of Boolean expression) ratio of NEQR approach is 1.5 times more than FRQI approach. The comparison of NEQR approach with other approaches is explained in Table 2.

**NAQSS Image representation approach [52]**
NAQSS is a multi-dimensional color image representation approach where (n+1) superpositioned qubits are used to represent a color image. In this technique, n qubits represent color and coordinates of $2^n$ pixels and 1 qubit represents an image segmentation information. All the pixels of an image have its corresponding value in the interval $[0,\pi/2]$ and a bijective function is used to represent one to one relationship between color and angle of those pixels. The NAQSS quantum image is represented as:

$$|I\rangle = \sum_{i=0}^{2^n-1} \theta_i |v_1\rangle|v_2\rangle \ldots \ldots |v_k\rangle \otimes |\chi_i\rangle \qquad (22)$$

Where, $|\chi_i\rangle = \cos\gamma_i |0\rangle + \sin\gamma_i |1\rangle$, where $\gamma_i$ corresponds to m to set up a bijective function if we divide an image into m sub-images that contains the pixel corresponding to the coordinate $|v_1\rangle|v_2\rangle \ldots \ldots |v_k\rangle$.

**SQR Image Representation [53]**

In SQR model, radiation energy of color objects is transformed into a quantum state |I> first. This conversion from energy state to the quantum state |I> is done by a special type of converter. This quantum state |I> composed of a set of qubit states $|\Phi_{ij}>$. A 2D detector array takes the responsibility to distribute the quantum states to the assigned position. So, SQR can be represented as,

$$|I> = |\Phi_{ij}>, i=1,2,.....N_1 \text{ and } j=1,2,.....N_2, \text{ where } |\Phi_{ij}> = \cos\theta_{ij}|0> + \sin\theta_{ij}|1> \tag{23}$$

A 2X2 quantum image can be represented in SQR mode as,

$$|I> = \{|\Phi_{00}>|\Phi_{01}>|\Phi_{10}>|\Phi_{11}>\}$$

Where $|\Phi_{00}> = \cos\theta_{00}|0> + \sin\theta_{00}|1>$, $|\Phi_{01}> = \cos\theta_{01}|0> + \sin\theta_{01}|1>$, $|\Phi_{10}> = \cos\theta_{10}|0> + \sin\theta_{10}|1>$ and $|\Phi_{11}> = \cos\theta_{11}|0> + \sin\theta_{11}|1>$.

**RGB Quantum Color Image Storage Scheme in Ternary Quantum System [55]:**
A new scheme has been proposed in the paper [54] where a RGB color image can be represented as 24 bit of color values and 2 bits of position values (in case of 3X3 color image). This allows storing an image with $N = 3^n \times 3^n$ pixels. As because a 3-level quantum register can hold maximum of $3^n$ basis states (by superposition), it needs total 33 basis states to represent a RGB color image. The storage circuit is built using basic MS gate, permutative gates and ternary Toffoli gate. Two famous techniques like FRQI and Normalized amplitude are modified and presented in this paper. This approach requires (m+n) qubits to store $2^m \times 2^n$ dimension real type of image, whereas FRQI approach requires (2m+1) qubits to store 2mX2m dimension real type of image. So, FRQI approach is suitable for square images but the proposed approach is suitable for all types of images. Also, if the numbers of qubits are increased then the quantum states are more prone to decoherence.

**Quantum image compression (QIC):** The main aim of Quantum image compression is to reduce the quantum resources (simple quantum gates) that are used to prepare quantum images. Compression means redundancy in the image can be reduced. Redundancy means repetition of the same information in the image. The reduction in the amount of simple quantum gates used for preparing the FRQI depends mainly on the reduction of the number of controlled rotation gates. To reduce the numbers of controlled rotation gates, a method is proposed which integrate controlled rotation gates which have the same rotation angle. The rotation angle and binary strings encoding conditional parts of rotation gates in a group characterize the group. From this point of view, each group has a universal controlled rotation gate in which the rotation angle is the group's rotation angle and the controlled condition is the amalgamation of all binary strings in the group. So these binary strings can be converted into its corresponding Boolean minterm form, where the corresponding Boolean minterm expression is minimized.This paper proposed a technique where quantum image can be compressed by applying minimization strategy on this transformed Boolean minterm form. As for example [24],
Suppose there are 4 positions (0, 8, 16, 32) having blue color in an 8×8 quantum image.

$$|0> = |000000> = \overline{x_5}\,\overline{x_4}\,\overline{x_3}\,\overline{x_2}\,\overline{x_1}\,\overline{x_0}$$
$$|8> = |001000> = \overline{x_5}\,\overline{x_4}\,x_3\,\overline{x_2}\,\overline{x_1}\,\overline{x_0}$$
$$|16> = |010000> = \overline{x_5}\,x_4\,\overline{x_3}\,\overline{x_2}\,\overline{x_1}\,\overline{x_0}$$
$$|32> = |100000> = x_5\,\overline{x_4}\,\overline{x_3}\,\overline{x_2}\,\overline{x_1}\,\overline{x_0}$$

Boolean expression
$e = \overline{x_5}\,\overline{x_4}\,\overline{x_3}\,\overline{x_2}\,\overline{x_1}\,\overline{x_0} + ................... + x_5\,\overline{x_4}\,\overline{x_3}\,\overline{x_2}\,\overline{x_1}\,\overline{x_0}$
So, Minimized expression,
$e_{min} = \overline{x_2}\,\overline{x_1}\,\overline{x_0}$

The Boolean expression captures all information about the binary strings in the group.This observation suggests that only one controlled-rotation gate can be used instead of 4 gates.So from here it can be concluded that the minimized Boolean expressions are used to construct a quantum circuit with a lesser number of simple gates than the original circuit[40].

**2.3 Applications of Quantum Natural computing into Image processing**

In this section, we present a short discussion on some of the important aspects of quantum image processing, such as quantum noise restoration, image compression based on quantum neural network approach, image and video processing applications on quantum computers, automatic object extraction using quantum computing. Image noise is random variation of brightness or color information in images and is usually a part of electronic noise. Image noise is an undesirable by product of image capture that adds false and irrelevant information [20].An image degradation happens due to linear frequency distortion and additive noise addition [29]. It develops a distortion measure(DM) of the effect of frequency distortion, and a noise quality measure(NQM) of the effect of additive noise [29].Till now the above discussion is circulated about classical noisy images but here it actually deals with the impact of quantum noise on images and how to remove it.According to paper[15],an approximate non-linear neural filter (NF) which is skilled to remove quantum noise from medical and natural images is introduced. Quantum noise (Poisson-distributed noise) is signal-dependent noise and is generally observed in photon-limited images such as infrared camera,X-ray images, and so on. The NF consists of a multilayer NN in which the activation functions of the units in the input, hidden and output layers are an identity function,a sigmoid function, and a linear function, respectively. Here the NF is trained by the renowned backpropagation algorithm to remove quantum noise from the image.This noise would depend on the average gray level in the input region. The input for all input units is calculated by making the approximate local average of all gray levels, except a certain variable target-input unit.Now the square-root operation is applied to a signal with Poisson noise.Then a signal with a constant noise variance (signal-independent additive noise) is obtained.Using this operation prior to filtering, it also may be possible to obtain a simpler implementation of the approximate filter for medical images. Thus the proposed approach is able to analyze unknown nonlinear deterministic systems with plural inputs such as the trained NFs [15].

This quantum noise can be also reduced by using some certain special transforms, like Anscombe Transform and Wiener Filter [16]. By reducing quantum noise a new image restoration method has been proposed here[16] and applied mainly on mammographic images. Anscombe Transform transforms quantum noise into additive noise. This procedure enables the use of any classical noise reduction technique, as Wiener filter, to reduce mammography image noise. The inverse filter was calculated based on the image system MTF, and was used after noise reduction procedure [16]. For realizing image compression and reconstruction, a three layer quantum backpropagation neural network based image compression approach is proposed in [17]. Genetic algorithm(GA) is used here to optimize neural network weights. Then combine the clamping based GA with quantum neural networks to finish image compression. Image segmentation is one of the vital low-level image processing task which is widely used in computer vision, pattern recognition etc. Quantum computing also plays a vital role to perform this task more efficiently and effectively. A paper [30] mainly deals with new image segmentation as a multiobjective optimization problem in which the aim is to get a set of non-dominated solutions, where quantum computing and evolutionary algorithms go hand by hand. The total approach divided into two phases: A split procedure using the well-known k-means algorithm followed by a merge procedure that uses a quantum-inspired evolutionary algorithm for the establishment of the non-dominated solutions set [30]. According to the approach, using k-means algorithm the original image is divided into k clusters, where each is composed of a set of pixels whose colours are close together. Then a small number of regions are filtered out with less than 10 pixels constraint. After that a multiobjective quantum inspired evolutionary algorithm is applied on resulting image partitioning to establish a set of non-dominated segmentations through integration of some adjacent regions. The algorithm uses only three quantum chromosomes. Each chromosome is a string of $N$ qubits where each qubit represents the edge separating two among the initial regions. The basic state $|0\rangle$ signifies that the edge is removed and the basic state $|1\rangle$ signifies that the edge is maintained. After generating the initial population set from that non-dominated set, four operations applied cyclically. They are:
- Quantum interference
- Quantum mutation
- Measurement
- Evaluation and update of the non-dominated solutions set

The evaluation is classified on two criteria: the intra-region homogeneity and the inter-region heterogeneity [30].

According to the paper [28], segmentation problem can be efficiently and accurately modeled by the quantum mechanics.This approach is mainly used for automatic object extraction in this paper [28]. This powerful segmentation method allows us to model complex objects and inherent structures of edge, shape and texture information along with the grey-level intensity uniformity, all in a single equation. The selection of the object segment is performed by maximizing a regularization energy function indicating the object boundaries [28].Besides the image segmentation,a famous quantum image edge detection algorithm using single qubit state and a quantum median filtering algorithm using dual qubits state are represented in the paper[9][23].They give their main focus on quantum measurements to retrieve output image from input image mapping.

The different versions of FRQI model are also used for various operations on images. A novel enhanced FRQI based image representation model is used for classification of images. This classification of digital images can be done using principal component analysis (PCA) and von Neuman quantum measurements. In this work, a PCA based classifier is used to detect images similar from the set of training images. Then the image signal space is divided into two orthogonal subspaces, one represents leading principal components and the other represents mostly noisy components. After that, the leading principal components are used to create a projector onto a subspace of quantum states. The image which is being classified is also encoded on a quantum state, and then measured using the projector defined [33].

A watermarking authentication scheme for quantum images is proposed based on the quantum cosine transform (QCT) and as an application of FRQI image representation model [34]. This proposed scheme utilizes a dynamic vector to control embedding strength instead of a fixed parameter for the embedding process in other schemes. The vector is an optimal solution of an optimization equation which is built to make the watermarked image having better visual quality. In this process, the FRQI based carrier image (|C>) sized $2^n \times 2^n$ is represented in its watermarking form, like

$$|W> = 1/2^n \sum_{i=0}^{2^{2n}-1} (\cos\theta_i |0> + \sin\theta_i |1>) \otimes |i> = \sum_{i=0}^{2^{2n}-1} |W_i> \otimes |i> \quad (24)$$

Now, quantum cosine transform (QCT) is applied on the carrier image(|C>) of sized $2^n \times 2^n$ and the quantum watermark image |W> is processed by some unitary transforms, like, 2-D identity matrix and a phase gate(|PW>).Now the quantum watermark image |W> is embedded into the QCT coefficients, like QCT(|CW>). After that, two registers are needed to encode the carrier image and the watermark image. Then, QFT operation is executed on the quantum carrier image |C> and a sequence of phase gates Ph(α)=P decided by embedding strength α is implemented on the watermark image. Finally,the adder quantum network introduced is implemented on the carrier image and the watermark image(Where *α* is the watermark embedding strength, it effects the quality of image after embedded).Finally, Executing the inverse QCT on QCT(|CW>),the watermarked quantum image |CW> is finally obtained [34].

## 3. Research Issues

Quantum image processing is an emerging field of research. Quantum algorithms can be used to solve problems defined within the realms of classical or quantum information. However, quantum algorithms can be developed to solve hard instances of the 3SAT Different technologies of quantum image processing are discussed so far. So, QIMP is an important field of open challenging problems for physicists, mathematicians, computer scientists, and engineers. Among all these ideas, some scopes for future work are focused in this section. They are briefly discussed as follows,

- According to paper [24],the approach is bounded by the main 3 types of image processing operators are used based on the concept of quantum Fourier transforms.The investigation on each of three types of image processing operations can also be applied on the quantum Wavelet transform,the quantum discrete cosine transform, etc. The mentioned approach directions are all on a single image. There are interesting questions on quantum operations having impacts on multiple images such as image matching, image searching on a set of images in FRQI states. These directions may open new results on quantum image processing in general.
- According to paper [15], in future such a neural filter can be made which perform the mathematical analyses on the statistical properties of the quantum noise and on the removal of noise using NNs. Using this approach, various kinds of NN models such as radial basis function networks etc. can be handled in future.
- Quantum image cryptography is one of the areas where we can apply the different features of quantum computing to get more compressed image which is suitable for storing and transformation.
- We can combine the various data mining techniques with quantum computing methods to recognize different image patterns [46], computer vision [48], automatic surveillance and moreover. Quantum information and quantum computation algorithms are also used in morphological image processing suitable for medical image analysis [45].
- Quantum Noise reduction from medical images is a big issue now a days. It is generated due to improper distribution of photons (pixels) over the receptor plane and the quantum noise can be reduced by increasing the concentration of photons over the plane. We can use any special technique which can deal with the density of the pixels of an image and also apply some density distribution technique to reduce the quantum noise.
- To remove the disadvantages associated with various famous models described in[24][27][38], we will have to focus in such a model where we may use multi qubit states that need to separate pixels with different colors (RGB) in order to apply distinct transformations. It makes easier to design quantum circuits for image processing tasks.
- Quantum computing and quantum information processing may play a key role in military applications. As for example, quantum radar can be built using the concept of quantum image processing algorithm (QIMP) [44].
- In the fields of astrophysics [47], bioinformatics, and nanotechnology etc, QIMP contribute toward solving open challenges in image processing.
- Image processing can be exploited in a multivalued quantum system with more benefits than a binary valued quantum system. As a future research scope,one can represents images (specially color images) in a n-qutrit quantum system by a superposition of $3^n$ basis states, thus a quantum register of size n can hold $3^n$ values simultaneously, whereas an n-qubits register can only hold only $2^n$ values. So it actually provides more dimensions in the Hilbert space to store large size of color images efficiently compare to the binary system. In case of image representation, the state space is increased by a factor of $(3/2)^m$, since the Hilbert space of *m* qutrits has the same dimensionality as $m\log_2 3$ qubits. The detailed comparison among various image representation techniques are given in Table 2.

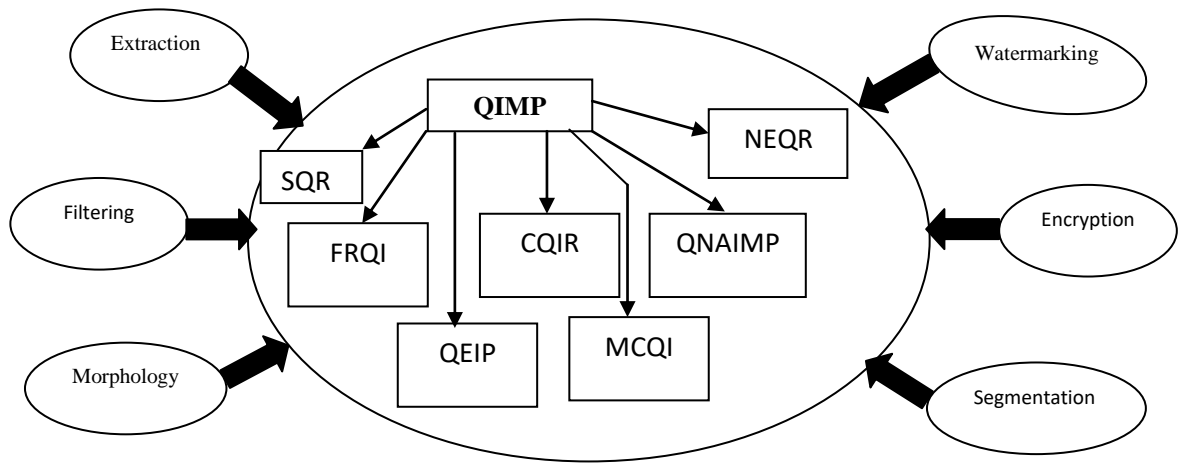

Fig.12 QIR algorithms and their applications

Table2. Comparison among various Quantum Image Representation Approaches

| FRQI Approach [24,40] | Normalized Amplitude based Approach [4] | Lattice representation model [43] | Image representation in Entangled quantum system [5, 22] | Image representation based on sequences of basis states approach[35, 41,42] | Novel enhanced quantum representation (NEQR) |
|---|---|---|---|---|---|
| The basis states of 2n qubit sequence is required to encode the position information of a pixel in a $2^n \times 2^n$ image and color information is encoded in the probability amplitudes of the corresponding one qubit state. According to the postulates of quantum mechanics, | The number of qubits increases with the image dimension while storing an image. So extra storage space is required to store information per pixel. One extra fractional bit is required to represent row-location vector or column-location vector. | Unlike FRQI, instead of storing the angle parameter of a color qubit, radiation energy value of each pixel is stored as a probability of projection measurement in this technique. | Less storage space is required due to the use of maximally entangled qubits which allows reconstructing images or shapes without using any additional /auxiliary information. | This model needs m qubits to represent $L=2^m$ color values. So unlike FRQI model (need one qubit) it occupies more memory space to represent color values of an image. It uses m qubit states rather than one qubit state to represent an image. | Unlike FRQI, NEQR approach can retrieve original image accurately through quantum measurement without using probability. However, more qubits are needed to encode a quantum image in NEQR representation. Such as, q+2n qubits are needed to construct a quantum |
| the probability amplitudes of a quantum state cannot be accurately defined using a finite number of measurements. Only simple color pixels operations are possible here due to the use of a single qubit state. | | | | | image model for a $2^n \times 2^n$ image with gray range $2^q$. (q qubits are needed to encode color information whereas 1 qubit required in FRQI representation). |
| Complex quantum circuit for implementation of this approach. | Less complex compare to FRQI model. | Less complex compare to FRQI model and simple quantum gates are used to build the circuit. | This method is only deployed for binary images. | Compared to the lattice representation model less quantum gates are required. | Less complex method compared to FRQI model. |
| The values of angel parameter are not quantified. So, it is difficult to identify which angle represents which color. Beside this, there is a practical limitation to physically represent angle parameter of a qubit. | No concept of angle parameters is here. | It can achieve a quadratic speedup in quantum image preparation but the quantification problem of angle parameter also exists in this model. | No concept of angle parameters is here. The entire process is using the concept of quantum entanglement. | The representable colors and positions of an image do not depend on the physical representation of angle parameters. | No concept of angle parameters is here. |
| - | - | The relationship between radiation | - | - | - |
|  |  | energy values and probabilities of projection measurement on qubits is quantified, which makes the representation model more flexible and clear. |  |  |  |
| This method is suitable for all sorts of images. | This method is valid for any color channel of any color space which represents an image, in part or whole. | Applications: infrared image operations. | Applications: Image segmentation, compression. | Applications: Image segmentation, filtering and more complex image processing operations can be applied. But this approach has been tested on grayscale images still now. | Image compression ratio is 1.5 times more than the FRQI approach. |

## 4. Conclusion

In order to achieve high performance image processing, quantum mechanics has been well explored to represent, store and process images. This article provides a brief survey of the current status of research in the quantum image processing paradigm. In this paper we have reviewed the principles regarding qubits formation and measurement of basis states, different issues of quantum image processing (image storage, representation, retrieval) and finally gives a brief description on the applications of this quantum image processing over various fields of computing. It discusses the different benefits of using quantum image processing over classical image processing. It also throws some light on the advantages of using multilevel quantum logic (qudit based) over qubit based quantum logic to represent a color image. The main purpose of this review is to gather the current mainstream of various Quantum image representations and discuss the advances made in the area. Besides that, some similarities, differences, shortcomings, and likely applications for some of the available Quantum image representation techniques are reviewed. The overall layout and structure of this paper is influenced by the contents of the papers [25][26]. All these efforts are needed for the realization of smooth, effective, fastest and secure QIMP technologies which help mankind to take full advantage of the immense potential of quantum computing and quantum information processing.